\documentstyle[12pt,epsf]{article}
\input epsfig
\newcommand{\postscript}[2]
{\setlength{\epsfxsize}{#2\hsize}
\centerline{\epsfbox{#1}}}
\newcommand{\nc}{\newcommand}
\nc{\pbarn}{\hbox {pb}}
\nc{\lumun}{\;{\hbox {pb}^{-1}}{\hbox {yr}^{-1}}}
\nc{\hc}{\hbox {h.c.}}
\nc{\re}{\hbox {Re}}
\nc{\im}{\hbox {Im}}
\nc{\mev}{\hbox {MeV}}
\nc{\gev}{\;\hbox {GeV}}
\nc{\etal}{\hbox{et al.}}
\nc{\prdj}[1]{{ \it Phys.~Rev.}~{\bf D{#1}}}
\nc{\prlj}[1]{{ \it Phys.~Rev.~Lett.}~{\bf {#1}}}
\nc{\plbj}[1]{{ \it Phys.~Lett.}~{\bf {#1B}}}
\nc{\npbj}[1]{{ \it Nucl.~Phys.}~{\bf B{#1}}}
\nc{\ptpj}[1]{{ \it Prog.~Theor.~Phys.}~{\bf {#1}}}
\nc{\zfpj}[1]{{ \it Z.~Phys.}~{\bf C{#1}}}
\nc{\mplaj}[1]{{ \it Mod.~Phys.~Lett.}~{\bf {#1A}}}
\nc{\beq}{\begin{equation}}
\nc{\eeq}{\end{equation}}
\nc{\bea}{\begin{eqnarray}}
\nc{\eea}{\end{eqnarray}}
\nc{\baa}{\begin{array}}
\nc{\eaa}{\end{array}}
\nc{\ra} {\rightarrow}
\nc{\ttbar}{t\bar{t}}
\nc{\bbbar}{b\bar{b}}
\nc{\tanb} {\tan \beta}
\nc{\twbdec} {t\rightarrow W^+ b}
\nc{\tbwbdec} {\bar{t} \rightarrow W^- \bar{b}}
\nc{\hprod} {e^+e^- \ra Z^\ast \ra H Z}
\nc{\epem} {e^+e^-}
\nc{\wpwm} {W^+W^-}
\nc{\tbar} {\bar{t}}
\nc{\bbar} {\bar{b}}
\nc{\wpp} {W^+}
\nc{\mt}{m_t}
\nc{\mts}{m_t^2}
\nc{\mj}{m_j}
\nc{\mjs}{m_j^2}
\nc{\mi}{m_i}
\nc{\mis}{m_i^2}
\nc{\mw} {m_W}
\nc{\mws} {m_W^2}
\nc{\mz} {m_Z}
\nc{\mzs} {m_Z^2}
\nc{\mh} {m_H}
\nc{\mhs} {m_H^2}
\nc{\hdec}{H \ra t\bar{t}}
\nc{\ttbardec}{\ttbar \ra W^+W^-\bbbar}
\nc{\wwbb}{W^+W^-\bbbar}
\nc{\po}{\Phi_1}
\nc{\pod}{\Phi_1^\dagger}
\nc{\pht}{\Phi_2}
\nc{\phtd}{\Phi_2^\dagger}
\nc{\phtt}{{\tilde{\Phi}}_2}
\nc{\popo}{\po^\dagger\po}
\nc{\phtpt}{\pht^\dagger\pht}
\nc{\popt}{\po^\dagger\pht}
\nc{\phtpo}{\pht^\dagger\po}
\nc{\as}{{\cal A}_{tt}}
\nc{\bt}{\beta_t}
\nc{\bts}{\beta_t^2}
\nc{\bz}{\beta_Z}
\nc{\bzs}{\beta_Z^2}
\nc{\bw}{\beta_W}
\nc{\bws}{\beta_W^2}
\nc{\sq}{\sqrt{2}}
\nc{\wphel}{\mbox{$\lambda_{W^+}$}}
\nc{\wmhel}{\mbox{$\lambda_{W^-}$}}
\nc{\bhel} {\mbox{$h_{b}$}}
\nc{\thel} {\mbox{$h_{t}$}}
\nc{\nsd} {N_{SD}}
\nc{\ntt} {N_{tt}}
\nc{\ashel}{{\cal A}_{\wphel,\wmhel}}

\begin{document}
%\begin{flushright}
%Version: \today\hfill
%\end{flushright}
\vspace*{2cm}
\begin{center}
{\large{\bf CP Violation in  {\boldmath $H \ra t \overline{t}$} Decays at
{\boldmath $\epem$} Colliders}\footnote{Supported in part by the Committee for
Scientific Research under grant 2 0165 91 01, Poland.}\\
\vspace*{2cm}
Bohdan Grz\c{a}dkowski\footnote{E-mail:{\tt bohdang@fuw.edu.pl.}}\\
\vspace{1cm}
        Institute for Theoretical Physics\\
        University of Warsaw\\ \vspace{1mm}
        Ho\.{z}a 69, PL-00-681 Warsaw, Poland}\\
\vspace*{3cm}

{\bf Abstract}
\end{center}
\vspace{5mm}
A helicity asymmetry for top quarks
originating from Higgs boson decays
is investigated within the 2-Higgs Doublet Model.
The asymmetry is sensitive  to CP
violation in the scalar sector of the model, whereas it vanishes in
the Standard Model.
It has been checked that without any fine tuning of parameters the
asymmetry can reach $50\%$. Standard decay patterns $\twbdec$ and
$\tbwbdec$ are utilized as spin analyzers to measure the asymmetry.
The Higgs production mechanism considered here is
$\hprod$. It has been shown that signal from the asymmetry can easy
overcome the noise
for Higgs bosons produced in future linear $\epem$ colliders at energy
$\sqrt{s}=500\gev$ operating with integrated luminosity
$L=2.0\times10^4\lumun$.

\vspace*{2.0cm}
\begin{flushright}
\parbox{2.in}
{IFT 07/94\\
hep-ph/9404330\\
March 1994}
\end{flushright}
\setcounter{page}{0}
\thispagestyle{empty}
\newpage

\section{Introduction}
In spite of spectacular successes of experimental high-energy
physics (e.g. precision tests of the Standard Model)
the origin of CP violation is still a mystery from both experimental
and theoretical point of view. As it is known very well there is only
one solid experimental signal of CP violation,
namely $K_L^0\ra\pi^+\pi^-$ decay~\cite{expcp}.
The classical method for incorporating CP violation into a model
of electroweak interactions has been proposed two decades ago
by Kobayashi and Maskawa~\cite{km}, it relays simply on explicit CP
violation through complex Yukawa couplings. This is the mechanism adopted
by the Standard Model (SM) of electroweak interactions. It is
obvious from the above that
our present experimental knowledge is very limited and theoretical tools
adopted as modest as possible. However, fortunately the simplest and
most attractive extension of the SM, 2-Higgs Doublet Model (2HDM)
provides much richer middles to
describe CP violation, namely spontaneous and/or explicite CP violation
in the scalar sector of the model~\cite{weinberg}. As will be seen in
the next section both scenarios result in mixing between imaginary and real
parts of complex Higgs fields, and a consequence of this is that mass
eigenstates
do not possess defined CP properties : CP must be violated! This paper is
devoted
to investigate possibilities for a detection of such sources of CP violation.

Linear high-energy $\epem$ collider (NLC) can prove to be
very useful laboratories
to study the physics of the Higgs boson. One of the many interesting issues
one can investigate there is, for instance, whether or not
CP is violated in $\hdec$  decay.
In this paper, some kind of helicity asymmetries in the final $\ttbar$
state will be discussed. In general, because of hadronization effects
such observables might be contaminated by large theoretical uncertainties.
The top quark however, because of its huge mass
offers few relevant advantages for the study of CP violation:
\begin{itemize}
\item{If $m_t>130$ GeV, it would decay before it can form a bound
state~\cite{bigi}; therefore the perturbative description is much more
reliable.}
\item{For the same reason, the spin information of the top quark would not be
diluted by hadronization, therefore helicity asymmetries can provide
a very useful tool in searching for CP violation.}
\item{Again, because of the large mass of the top quark, its properties are
sensitive to interactions mediated by Higgs bosons.}
\item{The Kobayashi-Maskawa~\cite{km} mechanism of CP violation
is strongly suppressed for majority of top-quark interactions;
therefore it is sensitive to non-conventional sources
of CP violation.}
\end{itemize}

There exists in the literature a number of papers~\cite{cpviol,keung}
investigating
CP violation within 2HDM in $\ttbar$ production,
Higgs boson decay and the top quark decay.
Here we will consider Higgs boson
production in $\epem$ future colliders through the classical Bjorken
mechanism $\hprod$ followed by the decay $\hdec$ where $\ttbar$
decays in a standard pattern: $\ttbardec$. The aim of this paper
was not only to include a realistic production mechanism into
account, but {\it first of all to establish a method of measurement
the asymmetry defined for
the $\hdec$ decay at the level of $\ttbar$ decay products}. We emphasize
that the method adopted here controls possible CP-violating effects
appearing in the process of Higgs boson production and also in
$\ttbar$ decays. Since the asymmetry will be measured at the level
of $\wwbb$ those effects may mix with CP violation in $\hdec$ and
influence an interpretation of measurements.
We shell prove that the asymmetry discussed here is free of those effects
and could be efficiently measured at future linear $\epem$ colliders.
\section{The Model}
We will consider here 2HDM of electroweak interactions. The model is
defined by Yukawa couplings and the Higgs potential.
It is known~\cite{wg} that
if two independent vacuum expectation values contributed to quark
mass matrices a model would predict tree-level flavour changing
neutral currents coupled to Higgs bosons. A standard solution to eliminate
them is to impose a discrete symmetry D:
\beq
\pht\ra-\pht,\;\;\;\;\;{u_i}_R\ra-{u_i}_R.
\label{dsym}
\eeq
Invariant quark Yukawa interactions read:
\beq
{\cal L}_Y=-(\bar{u}_i,\bar{d}_i)_L \Gamma_u^{ij} \phtt {u_j}_R
           -(\bar{u}_i,\bar{d}_i)_L \Gamma_d^{ij} \po  {d_j}_R + \hc,
\label{yukcoupl}
\eeq
where $i,j$ are generation indices and $\phtt$ is defined as
$i\sigma_2 \pht^\ast$.

The most general potential for the model is the following:
\bea
V(\po,\pht)&=&\delta V_{symm}(\po,\pht)+\delta V_{soft}(\po,\pht)+
\delta V_{hard}(\po,\pht) \\
\label{potential}
\delta V_{symm}(\po,\pht)&=&-\mu_1^2\popo-\mu_2^2\phtpt \nonumber  \\
&&+\lambda_1(\popo)^2+\lambda_2(\phtpt)^2+\lambda_3(\popo)(\phtpt) \nonumber \\
&&+\lambda_4|\popt|^2+\frac{1}{2}
\left[\lambda_5(\popt)^2+\hc\right] \nonumber \\
\delta V_{soft}(\po,\pht)&=&-\mu_{12}^2\popt+\hc  \nonumber \\
\delta V_{hard}(\po,\pht)&=& \frac{1}{2}\lambda_6|\popo|^2(\popt)+
\frac{1}{2}\lambda_7|\phtpt|^2(\popt)+\hc. \nonumber
\eea
The first term $\delta V_{symm}(\po,\pht)$ is symmetric under D symmetry.
It has been noticed~\cite{buras} long time ago that if one being orthodox
and na\"{\i}ve,
restricts himself
to absolutely symmetric Lagrangian (in order to keep a model renormalizable),
then the potential is just
$\delta V_{symm}(\po,\pht)$. However, a symmetric potential does not allow
neither for explicite (in the potential) nor spontaneous CP violation. That was
the reason why one had to introduce a third Higgs doublet in order to
remove FCNC and preserve spontaneous or explicite CP violation in a Higgs
sector. However, as showed by Symanzik~\cite{sym} soft breaking
of a symmetry (by operators dimension 3 and less) preserves renormalizability,
therefore it is allowed to add to the potential $\delta V_{soft}(\po,\pht)$
and as we shell see this is exactly what is needed to break
CP in the potential. Terms contained in $\delta V_{hard}(\po,\pht)$ break
the D symmetry hard and therefore can not by accepted.~\footnote{
In this case one would have to introduce all possible interactions of
dimension 4 to keep the model renormalizable. That means that FCNC
would be necessary.}

It is not difficult to find a condition on CP conservation; one can show
that if
\beq
\im({\mu_{12}^\ast}^4\lambda_5)=0
\label{cond}
\eeq
then there exist phases $\theta_{1,2}$ such that the following
CP transformation is a symmetry of the model:
\beq
\Phi_{1,2}\stackrel{CP}{\ra}\Phi_{1,2}^\dagger e^{i\theta_{1,2}}.
\label{cptrans}
\eeq
Therefore, now we know how to break CP explicitly in the potential.
It is easy to see that ``spontaneous'' (by noninvarainat vacuum
expectation values) CP breaking is also possible.
Let us first choose a phase of $\po$ field such that
$<\po>=v_1/\sqrt{2}$ is real
and a phase of $\pht$ such that $\lambda_5$ coupling is real, then one can
expect a complex vacuum expectation value for the second doublet
$<\pht>=v_2/\sqrt{2}e^{i\theta}$.
Solving minimum condition for
\beq
\lambda_5>0,\;\;\;\;\;\;\left|\frac{\mu_{12}^2}{2\lambda_5v_1v_2}\right|<1,
\eeq
we get~\cite{leephase}
\beq
\cos\theta=\frac{\mu_{12}^2}{2\lambda_5v_1v_2}.
\label{cpphase}
\eeq
So, we are able to conclude that in the 2HDM both explicit and spontaneous
CP violation is possible keeping the model renormalizable and avoiding FCNC
at the tree level. Two above scenarios lead to a mixing between real and
imaginary parts of Higgs fields in the mass matrix, a consequence of this
is that Yukawa interactions can not be invariant under CP:
\beq
{\cal L}_Y=h_i\bar{f}(a^f_i+ib^f_i\gamma_5)f,
\label{yukawy}
\eeq
where $h_i$ is a physical Higgs boson and $a^f_i$, $b^f_i$
are functions
of mixing angles which diagonalize the mass
matrix. In a basis, where longitudinal components of $Z$ decouple, the
rotation matrix is given by~\footnote{Notice a misprint in the formula
(21) of ref.~\cite{knowles}, the element (1,2) should read "$s_1c_3$".}
\beq
R=\left(\baa{ccc}
c_1&s_1c_3&s_1s_3\\
-s_1c_2&c_1c_2c_3-s_2s_3&c_1c_2s_3+s_2c_3\\
s_1s_2&-c_1s_2c_3-c_2s_3&-c_1s_2s_3+c_2c_3
\eaa\right),
\label{mixing}
\eeq
where $s_i\equiv\sin\alpha_i$ and $c_i\equiv\cos\alpha_i$. Parameters
of the Yukawa coupling (\ref{yukawy}) are given~\cite{knowles} in
terms of the elements
of $R$ matrix:
\bea
&a^u_i=-\frac{m_u}{v s_\beta}R_{i2}\;\;\;\;\;
b^u_i=-\frac{m_u}{v s_\beta}c_\beta R_{i3}&\nonumber\\
&a^d_i=-\frac{m_d}{v c_\beta}R_{i1}\;\;\;\;\;
b^d_i=-\frac{m_d}{v c_\beta}s_\beta R_{i3}&,
\label{abs}
\eea
where $v\equiv\sqrt{v_1^2+v_2^2}$, $s_\beta\equiv\sin\beta$,
$c_\beta\equiv\cos\beta$ and $\tan\beta\equiv v_2/v_1$.

\section{The Asymmetry}
In this section we will calculate the following CP violating helicity
asymmetry:
\beq
\as=\frac{\Gamma(++)-\Gamma(--)}{\Gamma(++)+\Gamma(++)},
\label{asymmetry}
\eeq
where by $\Gamma(++)/\Gamma(--)$ we understand a decay width of the lightest
Higgs boson
into $\ttbar$ pair with indicated helicities. Since under CP:
$(++)\leftrightarrow(--)$, nonzero $\as$ would be a signal of CP violation.
Let us define an effective
interaction of the lightest Higgs ($h_1$) with
a $\ttbar$ pair by:
\beq
{\cal L}_Y^{eff}=h_1\tbar(A+iB\gamma_5)t.
\label{effective}
\eeq
$A$ and $B$ are coefficients calculable order by order within a given model.
$\as$ can be expressed in terms of $A$ and $B$, at one-loop approximation
we get:
\beq
\as=\frac{-2\bt}{\bt^2{a^t_1}^2+{b^t_1}^2}\im[b^t_1A-a^t_1B].
\label{asres}
\eeq
Hereafter, for a particle of mass $m_i$, we adopt a notation
$\beta_i\equiv\sqrt{1-4m_i^2/m_1^2}$.
Let us define partial contributions $p_i$ to $\as$ through the formula:
\beq
\as=\frac{\sum_i p_i}{\bt^2{a^t_1}^2+{b^t_1}^2},
\label{hidef}
\eeq
where summation runs over all contributing diagrams shown in fig.~\ref{diag}.

\begin{figure}[t]  % t means top of the page, b- bottom, (..?)
\postscript{diagrams.ps}{0.7}              %**epsf**
\vspace{-0.25cm}
\caption{The diagrams contributing to the asymmetry $\as$.}
\label{diag}
\end{figure}

Let us list all partial contributions to the asymmetry.
Photon and gluon exchange give:
\beq
p_i=\alpha_i C_i a^t_1 b^t_1(1-\bt^2),
\label{agp}
\eeq
where i=QED, QCD, $C_{QCD}=4/3$ and $C_{QED}=4/9$. QCD contribution
is usually of the order of $10\%$, whereas the QED one is much smaller,
at the level of $10^{-1}\%$.

The single $Z$ exchange contribution can be written as:
\beq
p_Z=\frac{G_F a_1^t b_1^t\mts}{2 \sq \pi}\left[(1-\frac{8}{3}s_W^2)^2
\frac{\mzs}{m_1^2}F(\frac{m_1^2}{\mzs}\bts)+
F(\frac{m_1^2}{\mzs}\bts)(\bts+\frac{\mzs}{m_1^2})-2\bts\right],
\label{az}
\eeq
where $F(x)\equiv1-x^{-1}\log(1+x)$ and $s_W^2\equiv\sin^2\theta_W$.
Single $Z$ correction is always small, of the order of the QED one.

A Higgs boson exchange gives the following contribution:
\beq
p_{h_j}=\frac{1}{4\pi}(b_1^t a_j^t-\bts a_1^t b_j^t)(a_1^t a_j^t+
b_1^t b_j^t) F(\frac{m_1^2}{m_j^2}\bts),
\label{ah}
\eeq
where $j=1,2$ corresponds to two lightest Higgs bosons considered here.
Numerically, both $p_{h_1}$ and $p_{h_2}$ are very small, at the level of
$\sim 10^{-1}\%$ and $10^{-2}\%$, respectively (we use $m_2=430\gev$).

The $ZZ$-exchange contribution is the following:
\bea
p_{ZZ}&=&\frac{H^0_1}{32\pi}\frac{G_F}{\sq}m_1^2 b_1^t g \frac{\mt}{\mw}
\frac{\bz}{2\bt} \nonumber \\
&&\left\{P(\bt,\bz)\left[(1-\frac{8}{3}s_W^2)^2(1-\bzs)^2(1+2\bts+\bzs)
-4\bts(1+\bz^4)\right.\right.\\
&&+\left.\left.(1+\bzs)(3+\bz^4)\right]+
(1-\frac{8}{3}s_W^2)^2(1-\bzs)^2+(3+\bz^4)\right\},\nonumber
\eea
where the function $P$ is defined as:
\beq
P(x,y)\equiv\frac{1}{4xy}\log\left(\frac{1+x^2-2xy}{1+y^2+2xy}\right),
\label{pdef}
\eeq
$g$ is the $SU(2)$ gauge coupling constant and
$H^0_1=c_\beta c_1+s_\beta s_1 c_3$.

The $WW$ exchange diagram gives:
\bea
p_{WW}&=&\frac{H^0_1}{32\pi}\frac{G_F}{\sq}m_1^2 b_1^t g \frac{\mt}{\mw}
\frac{\bw}{\bt} \nonumber \\
&&\left[(2-L(\bw,\bt))\bts(3-\bws)+L(\bw,\bt)(1-\bws+2\bw^4)\right],
\eea
where
\beq
L(x,y)\equiv 1+\frac{x^2-y^2}{2xy}\log\left|\frac{x-y}{x+y}\right|.
\label{lfun}
\eeq
Both $WW$ and $ZZ$ diagram gives numerically relevant contributions
of the order of few per cent. The above contributions to the asymmetry
has been obtained before~\cite{keung}, we agree with their result, however
a sign convention adopted here is opposite.

Till now, in the existing literature~\cite{keung}, Higgs self-energy initial
state diagrams
have been
omitted. Usually their contributions are numerically small and sometimes
could be neglected.
However, there is a potentially large resonance effect possible, namely
if two lightest Higgs bosons have similar masses then mixed $h_1-h_2$
self-energy diagram gives a very large contribution, even if we stay not too
close to the resonance : $m_2-m_1>\Gamma_{1,2}$, where $\Gamma_{1,2}$ is a
width of the corresponding Higgs boson. The Higgs boson self-energy
contribution to the
asymmetry could be written as:
\beq
p_{12}=-2\bt(a_1^t b_2^t-b_1^t a_2^t)\im \frac{\Sigma_{12}}{m_1^2-m_2^2},
\label{ase}
\eeq
where contributions from $t$, $b$, $W$ and $Z$ to $\im \Sigma_{12}$ are listed
below
\bea
\im\Sigma_{12}^j&=&\frac{\bt N_C}{8\pi}m_1^2(a_1^j a_2^j\bts+b_1^j b_2^j)
\;\;\;\;j=t,b\;\;\;\;N_C=3\nonumber\\
\im\Sigma_{12}^W&=&\frac{\bw}{16\pi}\mws g^2 H^0_1 H^0_2
(3+\frac{m_1^4}{4\mw^4}\bws)\\
\im\Sigma_{12}^Z&=&\frac{\bz}{16\pi}\frac{\mzs}{c_W^2}g^2 H_1^0 H_2^0
(3+\frac{m_1^4}{4\mz^4}\bzs),\nonumber
\eea
where $H^0_2=-c_\beta s_1 c_2+s_\beta(c_1c_2c_3-s_2s_3)$.
The $h_1-Z$ self-energy gives the following contribution:
\beq
p_{hZ}=-\frac{G_F}{2 \sq \pi}a_1^t b_1^t \mts\bts
\label{ahz}
\eeq
After the asymmetry has been properly calculated we are
ready to rise a question:

\section{How to Measure the Asymmetry?}

Linear high-energy $\epem$ colliders provide a cleanest environment for the
Higgs
boson production. For mass ranges we are considering here the dominant
production mechanism is the Bjorken process: $\hprod$.~\footnote{
The production through the $WW$ or $ZZ$ fusion is less relevant
for $m_1>170\gev$ at $\sqrt{s}=500\gev$.}
After the Higgs boson is produced it will rapidly decay and we concentrate
on $\hdec$ mode here since we would like to measure the asymmetry $\as$.
However, $\ttbar$ pair would again decay and the only way to measure
$\as$ is to look for some observables defined at the level of $\ttbar$
decay products which would be sensitive to the asymmetry. Following ideas
from our previous paper~\cite{bg} we shell consider
here decays $\twbdec$ and $\tbwbdec$
and look for final states with defined $\wpwm$ helicities.

Let us consider top quark decays.
At the tree level in the high-energy
limit the $b$ quark always has helicity $\bhel=-$,
therefore helicity conservation
tells us that $W^+$ with helicity $\wphel=0$ coming from
the top of $\thel=+$ would like to go forward in the direction of
flying top, whereas $W^+$ with $\wphel=-$ will go mainly in the opposite
direction. $W^+$ emerging from the top of $\thel=-$ with the same
helicities as above would obviously go in the opposite direction.
An analogous picture holds for the $\bar{t}$ decays. It should be noticed
that the allowed helicities of $W^+$,
at the tree level and for massless $b$ quarks,
are $\wphel=0,-$,
and that helicity conservation never
permits to produce $\wphel=+/-$ together with $\bhel=-/+$,
respectively. Therefore, it seems natural to consider the following
asymmetries:
\begin{eqnarray}
{\cal A}_{\wphel \wmhel} & \equiv &
\frac{N_{\wphel \wmhel}}{D_{\wphel \wmhel}}\\
\label{definas}
N_{\wphel \wmhel} & \equiv & \left(
\int^{\frac{\pi}{2}}_0 d\theta \int^{\pi}_{\frac{\pi}{2}} d\bar{\theta}-
\int^{\pi}_{\frac{\pi}{2}} d\theta \int^{\frac{\pi}{2}}_0 d\bar{\theta}
\right)
\; \frac{d^2 \sigma_{\wphel \wmhel}}{d \theta\, d
\bar{\theta}}\\
\label{defn}
D_{\wphel \wmhel} & \equiv &
\int^{\pi}_0 d \theta \int^{\pi}_0 d \bar{\theta}
\; \frac{d^2 \sigma_{\wphel \wmhel}}{d \theta\, d \bar{\theta}},
\label{defd}
\end{eqnarray}
where $\theta,\,\bar{\theta}$ denote polar angles of $W^+,W^-$
measured in the $t,\bar{t}$ rest frames with respect to $t,\bar{t}$
directions seen from the Higgs boson rest frame,
respectively. $d^2\sigma_{\wphel \wmhel}/(d\theta\,d\bar{\theta})$
stands for the
cross-section obtained by integrating over
the full $\ttbar$ production phase space and over $W^+,W^-$ azimuthal
angles.

In order to calculate ${\cal A}_{\wphel \wmhel}$ we will need
helicity amplitudes for $\hdec$:
\bea
(++)&=&2E_t(b^t_1+ia^t_1 \bt)\nonumber\\
(--)&=&2E_t(b^t_1-ia^t_1 \bt)\\
(+-)&=&(-+)=0,\nonumber
\label{amp}
\eea
where $(h_t,h_{\bar{t}})$ denotes an amplitude for a given helicity and
$E_t$ is the top quark energy in the $\ttbar$ rest frame.

For top quark decays, $\twbdec$, the following
parameterization could be adopted:
\beq
\Gamma^\mu  =  \frac{-igV^{KM}_{tb}}{\surd 2}\bar{u}(p_b)\left[
\gamma^\mu (f^L_1 P_L+f^R_1 P_R)-\frac{i\sigma^{\mu \nu } k_\nu }{m_W}
(f^L_2 P_L+f^R_2 P_R)\right] u(p_t),
\eeq
where $P_{R/L}$ are projection operators, $k$ is the
$W$ momentum and $V^{KM}$ is the Kobayashi--Maskawa matrix.
Because $W$ is on shell,
two additional form factors do not contribute. At the tree level $f^L_1=1$ and
$f^R_1=f^L_2=f^R_2=0$.

Amplitudes for $W^+W^-$ production with helicities $(0,0)$
and $(-+)$ could be written
in terms of amplitudes for $\hdec$
$(h_t,h_{\bar{t}})$
and for $\twbdec$ $(h_t,\lambda_{W^+},h_b)$ and $\tbwbdec$
$\overline{(h_{\bar{t}},\lambda_{W^-},h_{\bar{b}})}$
decays, as follows:
\begin{eqnarray}
M_{00} & = &(++)(+0-)\overline{(+0+)}+(--)(-0-)\overline{(-0+)} \nonumber \\
M_{-+} & = &(++)(+--)\overline{(+++)}+(--)(---)\overline{(-++)}.
\label{amplitudes}
\end{eqnarray}
Since we are interested in the leading contribution to the asymmetries
(which is an interference between one-loop (see fig.~\ref{diag})
 and tree-level diagrams)
all amplitudes providing higher order corrections,
have been neglected above.

A direct calculation at the lowest order leads to the following result:
\beq
{\cal A}_{00}=-\frac{1}{2} \as\;\;\;\;\;{\cal A}_{-+}=+\frac{1}{2} \as
\label{asymm}
\eeq
It is very relevant to notice that, in the leading order, {\it there is no
contributions from CP violating effects in $\ttbar$ decays} ($\ttbar$ decays
enter at the tree level; $f^L_1=1,\;f^R_1=f^L_2=f^R_2=0$), therefore
the asymmetries ${\cal A}_{00/-+}$ measure directly CP violation
in the decay $\hdec$. It is worth to emphasize that possible CP
violating interactions in the production process $\hprod$ are irrelevant
since we observe inclusively all Higgs bosons produced
and therefore there
is no possibility to memorize anything concerning the production mechanism,
including possible CP violation. We can therefore conclude that
${\cal A}_{00/-+}$ are sensitive purely to CP violation in the Higgs
boson decay $\hdec$. There is a close analogy between asymmetries considered
here and those we have discussed in ref.~\cite{bg} for $\epem \ra \ttbar$.
However, there, we were also
able to find an asymmetry which was a good measure of CP violation
in $\ttbar$ decays. It was generated by $(0+)$ and $(-0)$
helicity configurations for $\wpwm$. However, to generate those configurations
one needs nonzero amplitudes $(+-)$ or $(-+)$ for the $\ttbar$ pair
what is, of course, impossible in our case, for $\ttbar$  pairs produced
in the decay $\hdec$.

It is very important to notice another consequence
of vanishing $(+-)$ and $(-+)$ amplitudes for $\ttbar$, namely,
in the case of Higgs decay
we {\it do not need to identify helicities of $\wpwm$ produced in the decay
process}. Following ref.~\cite{bg} it is useful to define the the asymmetry
${\cal A}_{tot}$,
where $(00)$ and $(-+)$ helicity states for $\wpwm$ are summed:
\beq
{\cal A}_{tot} \equiv \frac{N_{00}+N_{-+}}{D_{00}+D_{-+}},
\eeq
where $N_{\wphel \wmhel}$ are defined in the formula (\ref{defn}).
In the case of $\epem \ra \ttbar$ the above asymmetry has been introduced
in order to increase the statistics, however here, since $N_{0+}=N_{-0}=0$
the asymmetry counts all possible helicity configurations for $\wpwm$, and
therefore we do not need to measure $\wphel, \wmhel$.

It is easy to check that ${\cal A}_{tot}$ is directly related to our
initial asymmetry $\as$:
\beq
{\cal A}_{tot}= - \frac{1}{2}\;\;\frac{4\mw^4-\mt^4}{4\mw^4+\mt^4}\as.
\eeq
The above result makes the experimental determination of the asymmetry
much easier.\footnote{We thank W.-Y. Keung for bringing this point to
our attention.}

\begin{figure}[t]  % t means top of the page, b- bottom, (..?)
\postscript{asfixed.ps}{0.65}              %**epsf**
\vspace{-0.25cm}
\caption{The asymmetry $\as$ for $\tanb=0.5$ and various top quark masses;
$\mt=160,\; 170,\;180 \gev$ corresponding to solid, dashed and  dash-dotted
lines,
respectively. For the Higgs-boson mixing angles we have used
$\alpha_1=\pi/4$ and $\alpha_2=\alpha_3=3\pi/2$.}
\label{asfix}
\end{figure}

\section{Results}

According to most popular proposals for NLC we will consider
a machine operating at $\sqrt{s}=500\gev$. Since we are interested
in $\hdec$ decays, mass of the lighter Higgs boson will be varied
in the range $325-390\gev$, such that top quarks with mass between $160\gev$
and $180\gev$ could be produced. Since mass adopted for the next heavier
Higgs boson is $430\gev$ we are never closer to the propagator pole
(see $h_1-h_2$ self-energy graphs) as $40\gev$. A width of the lightest
Higgs boson $\Gamma_1$ varies in the range $4-39\gev$ (for $\tanb=.5$),
therefore
we never approach the pole at $m_2$ by more than $\Gamma_1$.
We assume here, that the charged Higgs boson is heavy:
$m_H>>\mw$. From $B^0-\bar{B}^0$ mixing and $\epsilon_K$ measurements
there exists a lower bound on $\tanb$~\cite{bgjg} as a
function of the charged Higgs boson mass $m_H$ and $\mt$,
however for heavy charged Higgs boson that bound is not very
restrictive and therefore we can safely consider $\tanb=.5$.

\newpage
\begin{figure}[t]  % t means top of the page, b- bottom, (..?)
\postscript{nsdmax.ps}{0.6}              %**epsf**
%\vspace{-0.25cm}
\caption{The maximal statistical significance $\nsd$ found for
indicated $\tanb$ and various top quark masses;
$\mt=160,\; 170,\;180 \gev$ corresponding to solid, dashed and dash-dotted
lines,
respectively. The curves terminate if the requirement to produce more than
$35$ $\ttbar$ pairs can not be satisfied.}
\label{nsdmax}
\end{figure}

In fig.~\ref{asfix} we present results for $\as$ obtained for fixed values of
the Higgs-boson mixing angles: $\alpha_1=\pi/4$, $\alpha_2=\alpha_3=3\pi/2$
and $\tanb=0.5$.
Hereafter we use
for QCD corrections running $\alpha_{QCD}$ evaluated at lighter Higgs mass.
As seen
from the figure the asymmetry could be substantial, at the level of $10\%$.
However, it is obvious that a big asymmetry is only a necessary condition to
observe it. Sufficiently large number of Higgs bosons must be produced
in order to see a signal. A necessary and sufficient condition which
must be satisfied to observe $\as$ at the $1-\sigma$ level is that
the asymmetry should be greater then statistical
fluctuations of the signal:
\beq
N_{SD}\equiv\sqrt{N_{tt}}\as>1,
\label{nsdef}
\eeq
where $N_{tt}$ is the number of $\ttbar$ pairs produced.

\newpage
\begin{figure}[t]  % t means top of the page, b- bottom, (..?)
\postscript{asmax.ps}{0.65}              %**epsf**
\vspace{-0.25cm}
\caption{An absolute value of the asymmetry $\as$
obtained for the parameter set which
maximize $N_{SD}$, for indicated $\tanb$ and various top quark masses;
$\mt=160,\; 170,\;180 \gev$ corresponding to solid, dashed and dash-dotted
lines,
respectively.}
\label{asmax}
\end{figure}

To calculate $\nsd$ we assume the Bjorken mechanism for the Higgs boson
production calculated consistently within 2HDM we are considering.
An integrated luminosity we are adopting is $L=2.0\times10^4\lumun$.
In order to find a maximal signal we scan over mixing angles
$\alpha_{1,2,3}$ for fixed value of $\tanb$, Higgs masses $m_1$, $m_2$ and the
top quark mass. Results for maximal $N_{SD}$ are presented in
fig.~\ref{nsdmax}.
It is seen that the $N_{SD}$ is substantial and the observation
should be possible. In fig.~\ref{asmax} we present values of the
asymmetry corresponding to maximal $N_{SD}$.
For $\tanb=.5$  $\as^{max}$ is at the level of $20-50\%$. Notice that since
for increasing $\tanb$ we suppress coupling to $\ttbar$,
therefore for $\tanb=1$ the maximal asymmetry is smaller.

\begin{figure}[t]  % t means top of the page, b- bottom, (..?)
\postscript{ascontrmax.ps}{0.65}              %**epsf**
\vspace{-0.25cm}
\caption{The gluonic(solid), $WW$ and $ZZ$ (dashed) and
$h_1-h_2$-self-energy(dotted)
contributions (absolute values) to the
asymmetry $\as$ corresponding to the parameter set which
maximize $N_{SD}$, for indicated top quark masses and $\tanb=.5$.}
\label{paref}
\end{figure}

The asymmetry
is a ratio of an appropriate rate difference to the total width
for the decay $\hdec$, therefore the maximalization procedure
is a competition between the minimal width and the maximal rate difference.
In order to make sure that we are not supperssing the width, we control
the total number of $\ttbar$ pairs produced and in the maximalization
procedure we always demand that at least 35 $\ttbar$ pairs must be
produced.

In order to illustrate the behavior of major effects we present in
fig.~\ref{paref} separated
contributions($\delta \as^{max}$) from
gluon, $WW$ and $ZZ$ exchange and mixed $h_1-h_2$ self energy
for $\tanb=.5$, again the numbers presented correspond to maximal $\nsd$.
As it could have been anticipated the gluonic contributions are decreasing
while
self-energy ones are increasing with Higgs masses since we are
approaching the pole at $m_1=m_2$.

Although, any detailed study of the background is beyond the scope of this
work, a few remarks are in order here. For the process $\hdec$ there is
of course a potential background coming from direct $\ttbar$ production,
however since we assume the Higgs mass known the invariant $\ttbar$ mass
cut should provide an effective way to remove balk of the background
assuming sufficient mass resolution.
The other useful tool is certainly a presence of a {\it monochromatic}
$Z$ boson in the
final state.
One should also have in mind that we do not take into account any
experimental cuts and, of course, some number of events
must be lost because of non-perfect efficiency.
The results that we have obtained here presumed the narrow width
approximation, where all possible interference effects between a
production and decays are neglected. In order to justify this
we must in addition
assume that the final $(Wb)$ mass resolution is sufficiently
good to be sure that $W$ and $b$ are coming from on-shell top quarks.

\section{Summary}

A helicity asymmetry $\as$ (see eq.(~\ref{asymmetry})) for top quarks
originating from Higgs boson decay
has been investigated within the 2HDM. The asymmetry is sensitive  to CP
violation in the Higgs sector of the 2HDM, whereas it vanishes in the SM.
We have checked that the
asymmetry can even reach $50\%$.
$\as$ is defined for the process $\hdec$, however, since top quarks
decay very fast the asymmetry can only be measured through $\ttbar$
decay products.

Standard decay patterns $\twbdec$ and $\tbwbdec$ has been
utilized~\footnote{Since we assume that the charged Higgs is very heavy,
the decay $t\ra Hb$ is kinematicaly forbidden.} as spin analyzers.
We have showed that an angular helicity asymmetries (see eq.~(\ref{definas}))
defined in terms of quantum numbers of $\ttbar$ decay products are simply
proportional to the $\as$ for the decay $\hdec$. It was important
to notice that the above proportionality holds even if we assume the
most general patterns for $\twbdec$ and $\tbwbdec$, what means that
$\ttbar$ decays
enter $\ashel$ only at the tree level and therefore can not provide any extra
source of CP noninvariance. We have checked that, in fact, one do not need
to measure helicities of $\wpwm$ and therefore {\it the angular asymmetry
${\cal A}_{tot}$ is a direct measure
of CP violation in $\hdec$.}

The Higgs production mechanism considered here was
$\hprod$. We have proved that signal from the asymmetry can easy
overcome the noise
for Higgs bosons produced in future linear $\epem$ colliders at energy
$\sqrt{s}=500\gev$ operating with integrated luminosity
$L=2.0\times10^4\lumun$. An experimental evidence for the asymmetry discussed
in this letter would be a direct signal of CP violation beyond the SM.

\vspace{1cm}
\centerline{\bf Acknowledgments}
\vspace{.5cm}

We thank J. Abraham for useful discussions.

\end{document}